\documentclass[aps,prl,superscriptaddress,reprint,floatfix]{revtex4-1}
\usepackage{graphicx,amssymb,amsmath,bm,dcolumn}
\usepackage[caption=false]{subfig}
\usepackage{natbib}
\usepackage{threeparttable}
\usepackage[bottom]{footmisc}
\usepackage{comment}
\usepackage{xfrac}

\usepackage{color}

\begin{document}

\renewcommand{\bibsection}{\section*{Bibliography}}

\title{Superfluid Optomechanics: Coupling of a Superfluid to a Superconducting Condensate}
\author{L. A. De Lorenzo}
\author{K. C. Schwab}
\affiliation{Applied Physics, California Institute of Technology, Pasadena, CA 91125 USA}
\email{schwab@caltech.edu}
\date{\today}

\begin{abstract}
We investigate the low loss acoustic motion of superfluid $^4$He parametrically coupled to a very low loss, superconducting Nb, TE$_{011}$ microwave resonator, forming a gram-scale, sideband resolved, optomechanical system.  We demonstrate the detection of a series of acoustic modes with quality factors as high as $7\cdot 10^6$.  At higher temperatures, the lowest dissipation modes are limited by an intrinsic three phonon process.  Acoustic quality factors approaching $10^{11}$ may be possible in isotopically purified samples at temperatures below 10 mK.   A system of this type may be utilized to study macroscopic quantized motion and as an ultra-sensitive sensor of extremely weak displacements and forces, such as continuous gravity wave sources.
\end{abstract}

\maketitle
\section{Introduction}

The study of the detection of motion at quantum mechanical limits has received great attention over the past 35 years\cite{braginsky1985,Schwab2005, Aspelmeyer2012} with much of the early thoughts and efforts focused on the engineering of very large-scale, ultra-sensitive gravitational wave antennas.  Given the recent success to prepare nano- and micron-scale mechanical structures at quantum limits,\cite{LaHaye2004,Teufel2011,Safavi2012} it is intriguing to consider what is required to accomplish quantum behavior with larger objects.

When considering such experiments, a key parameter which becomes apparent is the coupling rate to dissipative thermal environments which sets the timescale for energy decay, decoherence times\cite{zurek1992}, lifetimes of number states, limits of cooling to the ground state\cite{Roucheleau2009}, position sensitivity\cite{hertzberg2009}, and force sensitivity. Furthermore, fascinating experiments have recently been accomplished with small quantum condensates of atomic vapors coupled to optical resonators.\cite{Murch2008,Purdy2010}  In light of this, superfluid $^{4}$He, a condensate which can easily be prepared in macroscopic quantities and demonstrates frictionless motion at zero-frequency\cite{reppy1964}, is an intriguing material to consider for the study of quantized macroscopic motion\cite{Aspelmeyer2012} and quantized mechanical fields.\cite{Larraza1998}

Acoustic dissipation of first sound of superfluid $^4$He is a very well studied and understood process.\cite{chase1955, waters1967, abraham1968}  For temperatures below 0.5 K, the attenuation is due to non-linearities in the compressibility of liquid helium, which couples the low frequency acoustic mode with thermal phonons, and leads to an acoustic attenuation length of:\cite{abraham1968}
\begin{equation}
\alpha = \frac{\pi^3}{60} \frac{\left(G+1\right)^{2} }{\rho_{4} \hbar^3 c_{4}^{6}} \left(k_{B}T\right)^{4}\omega
\end{equation}
where $G=\left(\rho/c\right)\partial c / \partial \rho=2.84$ is the Gr\"{u}neisen's parameter,\cite{Abraham1970} $k_B$ is the Boltzmann constant, $\rho=145$ kg/m$^{3}$, $\hbar$ is the Planck constant, $c_4=238$ m/s is the speed of sound,\cite{Abraham1970} $\omega$ is the frequency of the acoustic wave, and $T$ is the temperature. Assuming a sample temperature of 10 mK this leads to an acoustic quality factor of $Q=5\cdot10^{10}$, which would exceed the highest recorded mechanical quality factors\cite{mcguigan1978, bagdasarov1977, goryachev2012} of $10^{9}$.  A 5 kHz acoustic mode in this condition would have an extraordinary number state lifetime of $\tau_N=\hbar Q/(k_B T)\approx 36$ s.

\begin{figure*}[t]
\begin{centering}
\includegraphics[width=.8\textwidth]{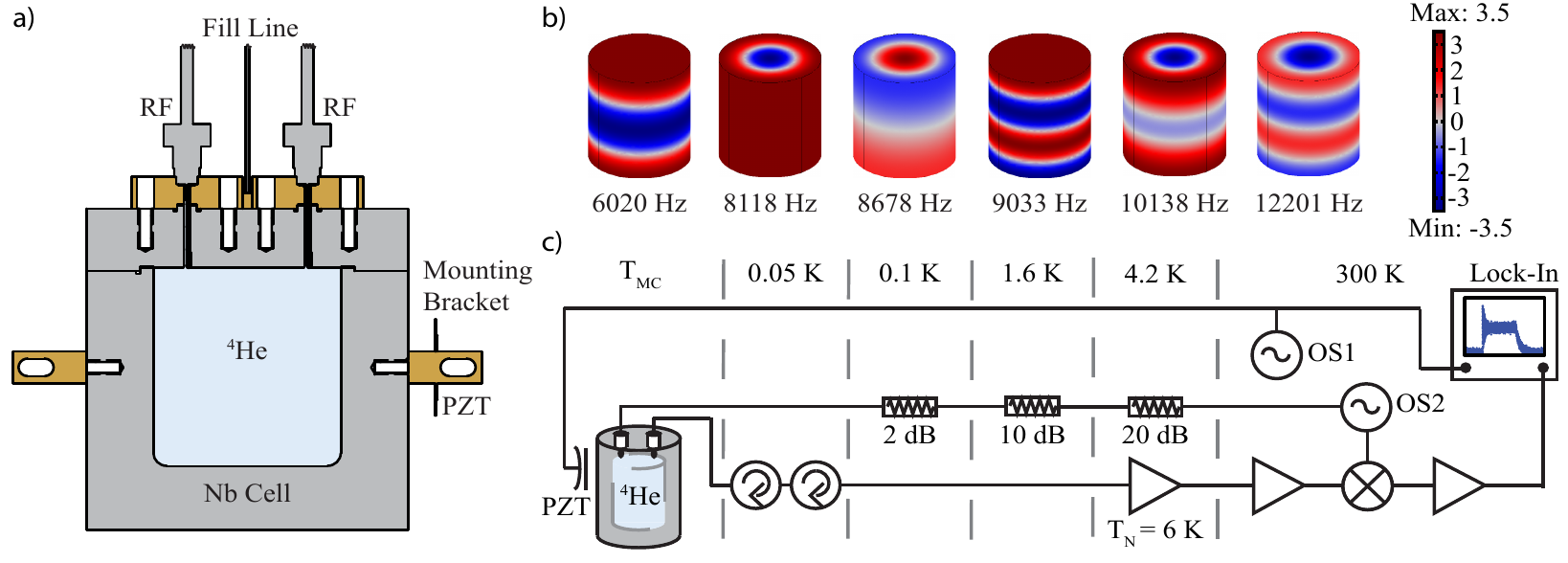} 
\end{centering}
\caption{a)  shows a cross-section of the parametric system: a niobium cavity filled with superfluid $^4$He.  The resonator is formed from two pieces (body and lid) with the microwave ports and helium fill-line located on the lid and sealed to the lid with indium.  The lid is sealed to the body with indium which has a $T_c$ of 3.4K.  The volume of the cavity is $39.3$ cm$^3$ and is filled with 5.70 grams of $^4$He.  b) shows the pressure profile of the lowest frequency superfluid acoustic resonances computed with finite element analysis; pressure nodes are shown in white. c) shows the low temperature microwave measurement circuit, where OS1 is at audio frequency and used to drive the piezoelectric actuator (PZT) which drives the acoustic mode, and OS2 is a microwave oscillator used to pump the SCR.}
\label{fig1}
\end{figure*} 
A parametric system is formed when low frequency mechanical motion modulates a higher frequency electromagnetic resonance: an optical cavity with a mechanical resonator as a mirror, or a microwave cavity coupled to a mechanical element are prototypical examples of parametric opto-mechanical systems.\cite{Aspelmeyer2012}  In this work, we couple the acoustic motion of superfluid  $^{4}$He to a microwave resonance through the modulation of the density and resulting modulation of the permittivity.  We find\cite{DeLorenzo2013} that the change in the resonant microwave frequency due to an acoustic wave with pressure amplitude P is:
\begin{equation}
\frac{\partial \omega_{C}}{\partial P} = - \frac{\omega_{C}  \kappa_{He}}{6} \left( \epsilon_{R} + 2 \right) \left(\epsilon_{R} -1 \right)  \Omega
\end{equation}  
where $\omega_C$ is the microwave cavity frequency, $\kappa_{He}=1.2\cdot 10^{-7}$ Pa$^{-1}$ and $\epsilon_R=1.057$ are the compressibility and relative permittivity of liquid helium.   $\Omega$ is the geometric coupling between the acoustic wave and the energy stored in the microwave field:
\begin{equation}
\Omega=\frac{ \int  f \left( r, \theta, z \right)   \left|E \left( r, \theta, z \right)\right|^{2} \,dV }{ \int  \epsilon_{R} \left|E\left( r, \theta, z \right)\right|^{2} \,dV }
\end{equation}
 where $P\cdot f\left(r,\theta,z\right)$ and $E \left( r, \theta, z \right)$ are the spatial functions of the pressure field of the acoustic mode and electric field of the microwave mode.    For the TE$_{011}$ mode coupled to the acoustic mode at 10138 Hz (see Fig. 1), we compute $\Omega = -.083$.

The coupling between the motion of the helium and the microwave field is relatively weak in comparison to nano- and micro-scale optomechanical realizations: the single quanta frequency shift is given by $\frac{\partial \omega_{C}}{\partial P} \Delta P_{SQL}=3.3\cdot 10^{-8}$ where  $\Delta P_{SQL}=\sqrt{\hbar \omega/(\kappa_{He} V_{eff})} =2\cdot 10^{-9}$ Pa is the amplitude of the zero point fluctuations of the acoustic field, $V_{eff}=19.5$ cm$^3$ is the effective volume of the acoustic mode and $\omega=2\pi\cdot 10$ kHz.\cite{DeLorenzo2013} However, as a result of the extremely low levels of dissipation of the microwave resonator and the very low dielectric loss of the helium ($\epsilon^{''}<10^{-10}$)\cite{hartung2006}, we do expect to be able to pump this system sufficiently hard to detect the acoustic field at the Standard Quantum Limit (SQL).\cite{DeLorenzo2013}   Furthermore, mechanical coupling between the superfluid and the Nb cell is also expected to be weak given the large difference in speeds of sound ($c_{Nb}= 3480$m/s): the lowest acoustic resonance of the helium is lower in frequency than the lowest vibrational mode of the Nb cell.  In addition, Nb has been shown to be an excellent mechanical material at low temperatures (Q of $4\cdot 10^7$ at 50 mK at kHz\cite{duffy1994}).  Finite element analysis shows that only $\approx4\cdot10^{-4}$ of the acoustic energy resides in the Nb cell with the superfluid excited at the 6020 Hz acoustic resonance (Fig. 1b,) limiting\cite{DeLorenzo2013} the quality factor of the acoustic helium mode to $10^{11}$.  Other designs utilizing very low loss dielectric materials such as sapphire are also possible and under development.\cite{DeLorenzo2013}

Our superconducting cavity resonator (SCR) is formed from a billet of superconductivity-grade Nb\cite{ATIWahChang} which is machined into cylindrical cavity with internal dimension: 1.78 cm radius and 3.95 cm length ( Fig 1a.)   After machining, the Nb pieces were polished and etched in acid (HF:HNO$_3$:HPO$_3$::1:1:2) for 40 min which removed approximately $100\mbox{ }\mu$m of material.\cite{Padamsee2009}

Microwaves are coupled through loops recessed into the lid.  Measurements at 1.7 K show a TE$_{011}$ resonance of 10.89 GHz (10.60 GHz when filled with He) with an internal loss rate of $\kappa_{int}=(2\pi)\cdot 31$ Hz.  After characterizing the SCR, we increased the coupling to $\kappa_{in}=\kappa_{out}=(2\pi)\cdot 633$ Hz, enabling sideband resolution for kilo-hertz acoustic modes.  The SCR is coupled to the mixing chamber of the dilution refrigerator (base temperature of 5.5mK) with two copper brackets.  

The cell is filled at 4.0 K through a 0.5 mm ID capillary with normal isotopic purity $^4$He. As the cell is cooled below 1K, the helium undergoes a substantial thermal contraction ($\Delta \rho/\rho=0.13$), which lowers the free surface of the liquid helium into a 7.4 cm$^3$ volume placed at the mixing chamber, before the SRC.  This avoids the heat load into both the helium and the refrigerator from a filled capillary.  The remaining heat load from the helium film in the capillary limits the base-temperature of the refrigerator to 33 mK.  The superfluid cools through the walls of the SRC with a thermal time constant expected to be ~$\sim 10^3$ seconds, which is anticipated to be largely temperature independent.\cite{Lounasmaa1974}  
Future work will include the addition of a low temperature valve\cite{Sprague1998} to the fill-line which will avoid both the heat flow and acoustic coupling into the superfluid cell.  
\begin{figure}[t]
\begin{centering}

\includegraphics[width=.9\linewidth]{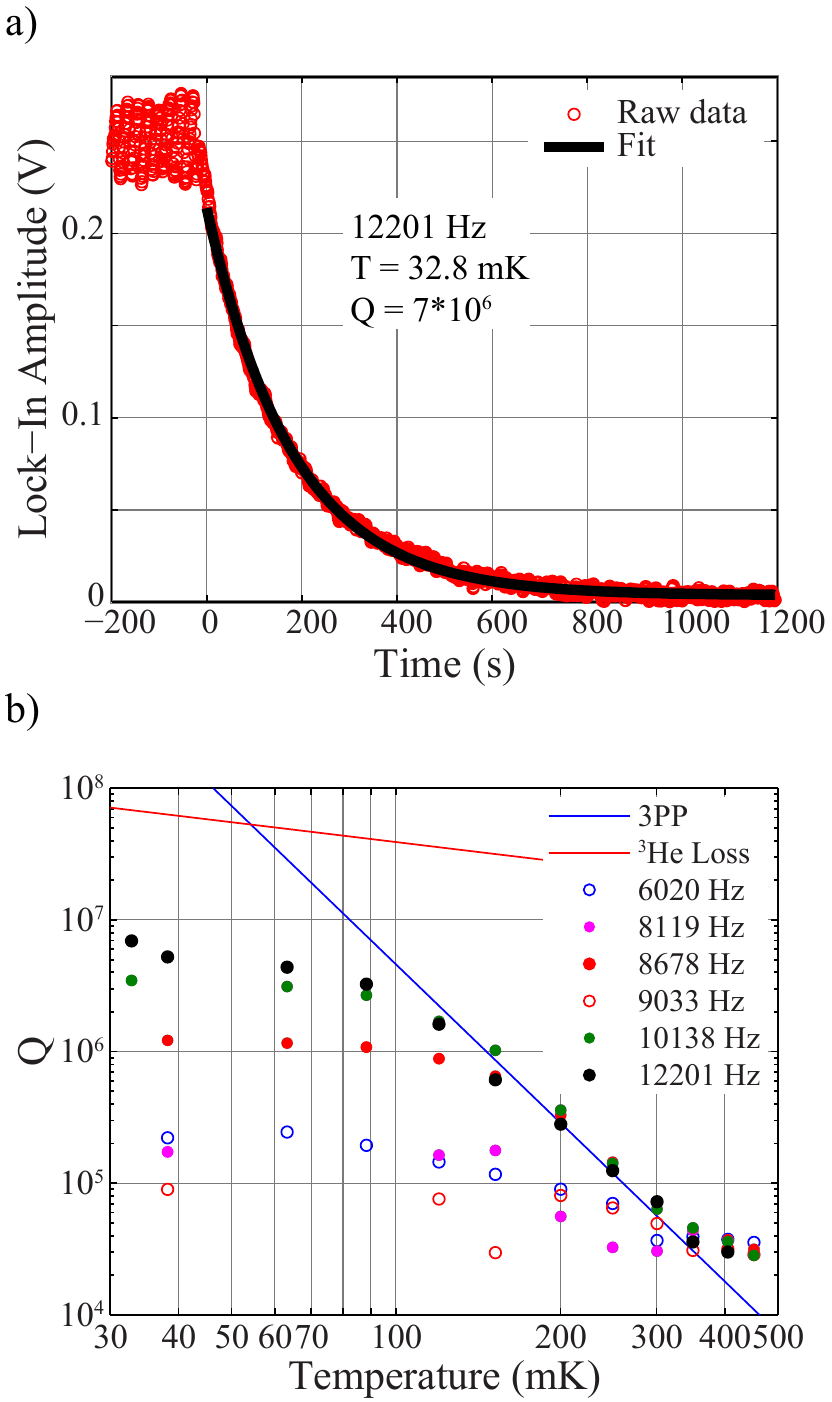} 
\end{centering}
\caption{a) shows the free decay of the helium acoustic resonance at 12,201 Hz, demonstrating a quality factor of $7\cdot 10^6$.   b) shows the measured quality factors for a number of the lowest frequency acoustic modes, versus refrigerator temperature, where modes with a radial node are shown as filled symbols, and modes without are shown is unfilled.  The blue line shows the dissipation expected from 3-phonon process (Eq.1), and the red line shows the dissipation expected from $^3$He impurities at a concentration of $10^{-6}$.}
\label{fig2}

\end{figure} 

We pump the parametric system with a microwave source, red-detuned from the SRC resonance and measure the up-converted sideband appearing at the cavity frequency\cite{Roucheleau2009} due to the acoustic motion which is stimulated with a piezoelectric transducer attached near the SRC.  We find a series of acoustic resonances within 1\% of the expected frequencies for a right cylindrical resonator. We measure the quality factor by recording the undriven, free decay.  Figure 2a show such a trace for a 12,201 Hz mode demonstrating a Q of $7\cdot 10^{6}$.  Figure 2b shows the measured quality factors for a number of the acoustic modes versus temperature. 

For the highest Q acoustic modes, the Q approximately follows the dissipation rate versus temperature due to the expected 3-phonon process (Eq.1), but eventually saturates below 100 mK.  At this point we do not have a clear understanding of this saturation or the difference in Q between the modes: from Eq. 1 we would expect the Q to be frequency independent.  It is interesting to note that the higher Q modes have a pressure node near the location of the fill line (radial node), suggesting that the acoustic loss into the fill line may be dominating the dissipation of lower Q resonances.  As the frequency approaches the first acoustic mode of the niobium cell (15 kHz), the helium acoustic resonances become more difficult to isolate, and the Q's start to degrade.  Above 15 kHz, some modes are extremely low Q ($Q<10^{3}$).  We have also estimated the dissipation due to $^3$He impurities\cite{DeLorenzo2013} and show this on Fig. 2a.  We believe the highest Q acoustic modes are not yet limited by this effect, which will scale as $\omega^{2}$ and is inconsistent with our data.  We have obtained a sample of isotopically purified helium\cite{mcclintock1978} (concentration of $^3$He is $10^{-10}$ that of $^4$He), to mitigate this effect in future experiments.

The measurement shown in Fig. 2a are realized with a microwave pump strength producing $n_p = 4.5\cdot10^{8}$ microwave photons inside the SRC, leading to a dissipation in the SRC of 0.6 pW, and in the helium itself of $<20$ fW.  At 100 mK (10 mK), this would lead to an increase of the temperature of the helium of 1 nK (1 $\mu$K).  This low level of heating allows for the application of large amplitude microwave signals until the sensitivity is limited by the phase noise of the microwave sources.  With our lowest phase noise source (Agilent E8257D-$\mathcal{L}\left( 10 kHz\right)=-110 db_c/Hz$) and a 10 kHz, $Q=3\cdot 10^6$ acoustic resonance, the pressure resolution is $3\cdot 10^{-3}$ Pa, which corresponds to a noise temperature of $1.5\cdot 10^6$ K.  Utilizing a sapphire cryogenic reference, a much lower phase noise source is possible: $\mathcal{L}\left( 10 kHz\right)=-156 db_c/Hz$\cite{Woode1996}.  This source combined with a Q=55M SRC as a filter cavity would yield a detection noise temperature of 10 mK with $n_P = 5\cdot10^{13}$.  Realizing an acoustic Q of $5\cdot10^{10}$ would allow for detection near the SQL (noise temperature of $2 \cdot 10^{-7}$ K) and sideband cooling of the acoustic mode.\cite{Roucheleau2009}

The possibility to produce an extremely sensitive inertial sensor can be understood through the following estimate.  Suppose the helium acoustic resonance is modeled as a simple mass-spring system contained inside the Nb cavity, where $m=5.7$ g is the mass of the superfluid and $k=m\omega^2=10^7$ N/m.  The thermal motion of this system at 100 mK is $x_{th}=\left(k_bT/(m\omega^2)\right)^2=2\cdot10^{-16}$ m.  If one drove this system by shaking the Nb cavity at the acoustic resonance, with an amplitude $x_0$, the motion of the helium would be $x=Q\cdot x_0$.  Turning this expression around, and assuming both our current device parameters (10 kHz, $Q=3\cdot10^6$, 100 mK) and that the detection is limited by the thermal noise of the acoustic field, we would expect a displacement sensitivity of the Nb container to be $8 \cdot 10^{-23}$ m; assuming 10 mK and $Q=10^{11}$, we expect sensitivity to motion of the outside container of $x_0=x_{th}/Q=8 \cdot 10^{-28}$ m.

This remarkable level of sensitivity would correspond to a strain sensitivity of $h\approx 2 \cdot 10^{-26}$ assuming our current device geometry.  This level of strain is below the level predicted for rapidly spinning pulsars.\cite{Abbott2008,Press1972}
 Although this extreme sensitivity would only be obtained around the narrow acoustic resonance, it is possible to tune the acoustic resonance by as much as 50\% by pressurizing and modifying the speed of sound of the helium.\cite{Atkins1953,Abraham1970}  This frequency control could also enable  parametric amplification techniques which could both preamplify weak signals and provide sensitive detection at frequencies other than the resonant acoustic frequency.  


\begin{acknowledgments}
We would like to acknowledge helpful conversations with Rana Adhikari. We acknowledge funding provided by the Institute for Quantum Information and Matter, an NSF Physics Frontiers Center(NSF IQIM-1125565) with support of the Gordon and Betty Moore Foundation (GBMF-1250) NSF DGE-1144469, NSF DMR-1052647, and DARPA-QUANTUM HR0011-10-1-0066.
\end{acknowledgments}

\bibliographystyle{apsrev4-1}
\bibliography{sf_res_bib_v21}

\begin{thebibliography}{32}%
\makeatletter
\providecommand \@ifxundefined [1]{%
 \@ifx{#1\undefined}
}%
\providecommand \@ifnum [1]{%
 \ifnum #1\expandafter \@firstoftwo
 \else \expandafter \@secondoftwo
 \fi
}%
\providecommand \@ifx [1]{%
 \ifx #1\expandafter \@firstoftwo
 \else \expandafter \@secondoftwo
 \fi
}%
\providecommand \natexlab [1]{#1}%
\providecommand \enquote  [1]{``#1''}%
\providecommand \bibnamefont  [1]{#1}%
\providecommand \bibfnamefont [1]{#1}%
\providecommand \citenamefont [1]{#1}%
\providecommand \href@noop [0]{\@secondoftwo}%
\providecommand \href [0]{\begingroup \@sanitize@url \@href}%
\providecommand \@href[1]{\@@startlink{#1}\@@href}%
\providecommand \@@href[1]{\endgroup#1\@@endlink}%
\providecommand \@sanitize@url [0]{\catcode `\\12\catcode `\$12\catcode
  `\&12\catcode `\#12\catcode `\^12\catcode `\_12\catcode `\%12\relax}%
\providecommand \@@startlink[1]{}%
\providecommand \@@endlink[0]{}%
\providecommand \url  [0]{\begingroup\@sanitize@url \@url }%
\providecommand \@url [1]{\endgroup\@href {#1}{\urlprefix }}%
\providecommand \urlprefix  [0]{URL }%
\providecommand \Eprint [0]{\href }%
\providecommand \doibase [0]{http://dx.doi.org/}%
\providecommand \selectlanguage [0]{\@gobble}%
\providecommand \bibinfo  [0]{\@secondoftwo}%
\providecommand \bibfield  [0]{\@secondoftwo}%
\providecommand \translation [1]{[#1]}%
\providecommand \BibitemOpen [0]{}%
\providecommand \bibitemStop [0]{}%
\providecommand \bibitemNoStop [0]{.\EOS\space}%
\providecommand \EOS [0]{\spacefactor3000\relax}%
\providecommand \BibitemShut  [1]{\csname bibitem#1\endcsname}%
\let\auto@bib@innerbib\@empty
\bibitem [{\citenamefont {Braginsky}\ \emph {et~al.}(1985)\citenamefont
  {Braginsky}, \citenamefont {Mitrofanov},\ and\ \citenamefont
  {Panov}}]{braginsky1985}%
  \BibitemOpen
  \bibfield  {author} {\bibinfo {author} {\bibfnamefont {V.~B.}\ \bibnamefont
  {Braginsky}}, \bibinfo {author} {\bibfnamefont {V.~P.}\ \bibnamefont
  {Mitrofanov}}, \ and\ \bibinfo {author} {\bibfnamefont {V.~I.}\ \bibnamefont
  {Panov}},\ }\href@noop {} {\emph {\bibinfo {title} {Systems with small
  dissipation}}}\ (\bibinfo  {publisher} {The University of Chicago Press},\
  \bibinfo {year} {1985})\BibitemShut {NoStop}%
\bibitem [{\citenamefont {Schwab}\ and\ \citenamefont
  {Roukes}(2005)}]{Schwab2005}%
  \BibitemOpen
  \bibfield  {author} {\bibinfo {author} {\bibfnamefont {K.}~\bibnamefont
  {Schwab}}\ and\ \bibinfo {author} {\bibfnamefont {M.}~\bibnamefont
  {Roukes}},\ }\href@noop {} {\bibfield  {journal} {\bibinfo  {journal}
  {Physics Today}\ }\textbf {\bibinfo {volume} {58}},\ \bibinfo {pages} {36}
  (\bibinfo {year} {2005})}\BibitemShut {NoStop}%
\bibitem [{\citenamefont {Aspelmeyer}\ \emph {et~al.}(2012)\citenamefont
  {Aspelmeyer}, \citenamefont {Meystre},\ and\ \citenamefont
  {Schwab}}]{Aspelmeyer2012}%
  \BibitemOpen
  \bibfield  {author} {\bibinfo {author} {\bibfnamefont {M.}~\bibnamefont
  {Aspelmeyer}}, \bibinfo {author} {\bibfnamefont {P.}~\bibnamefont {Meystre}},
  \ and\ \bibinfo {author} {\bibfnamefont {K.}~\bibnamefont {Schwab}},\
  }\href@noop {} {\bibfield  {journal} {\bibinfo  {journal} {Physics Today}\
  }\textbf {\bibinfo {volume} {65}},\ \bibinfo {pages} {29} (\bibinfo {year}
  {2012})}\BibitemShut {NoStop}%
\bibitem [{\citenamefont {LaHaye}\ \emph {et~al.}(2004)\citenamefont {LaHaye},
  \citenamefont {Buu}, \citenamefont {Camarota},\ and\ \citenamefont
  {Schwab}}]{LaHaye2004}%
  \BibitemOpen
  \bibfield  {author} {\bibinfo {author} {\bibfnamefont {M.~D.}\ \bibnamefont
  {LaHaye}}, \bibinfo {author} {\bibfnamefont {O.}~\bibnamefont {Buu}},
  \bibinfo {author} {\bibfnamefont {B.}~\bibnamefont {Camarota}}, \ and\
  \bibinfo {author} {\bibfnamefont {K.~C.}\ \bibnamefont {Schwab}},\
  }\href@noop {} {\bibfield  {journal} {\bibinfo  {journal} {Science}\ }\textbf
  {\bibinfo {volume} {304}},\ \bibinfo {pages} {74} (\bibinfo {year}
  {2004})}\BibitemShut {NoStop}%
\bibitem [{\citenamefont {Teufel}\ \emph {et~al.}(2011)\citenamefont {Teufel},
  \citenamefont {Donner}, \citenamefont {Li}, \citenamefont {Harlow},
  \citenamefont {Allman}, \citenamefont {Cicak}, \citenamefont {Whittaker},
  \citenamefont {Lehnert},\ and\ \citenamefont {Simmonds}}]{Teufel2011}%
  \BibitemOpen
  \bibfield  {author} {\bibinfo {author} {\bibfnamefont {J.}~\bibnamefont
  {Teufel}}, \bibinfo {author} {\bibfnamefont {T.}~\bibnamefont {Donner}},
  \bibinfo {author} {\bibfnamefont {D.}~\bibnamefont {Li}}, \bibinfo {author}
  {\bibfnamefont {J.}~\bibnamefont {Harlow}}, \bibinfo {author} {\bibfnamefont
  {M.}~\bibnamefont {Allman}}, \bibinfo {author} {\bibfnamefont
  {A.}~\bibnamefont {Cicak}, \bibfnamefont {K.~Sirois}}, \bibinfo {author}
  {\bibfnamefont {J.}~\bibnamefont {Whittaker}}, \bibinfo {author}
  {\bibfnamefont {K.}~\bibnamefont {Lehnert}}, \ and\ \bibinfo {author}
  {\bibfnamefont {R.}~\bibnamefont {Simmonds}},\ }\href@noop {} {\bibfield
  {journal} {\bibinfo  {journal} {Nature}\ }\textbf {\bibinfo {volume} {475}},\
  \bibinfo {pages} {359} (\bibinfo {year} {2011})}\BibitemShut {NoStop}%
\bibitem [{\citenamefont {Safavi-Naeini}\ \emph {et~al.}(2012)\citenamefont
  {Safavi-Naeini}, \citenamefont {Chan}, \citenamefont {Hill}, \citenamefont
  {Alegre}, \citenamefont {Krause},\ and\ \citenamefont
  {Painter}}]{Safavi2012}%
  \BibitemOpen
  \bibfield  {author} {\bibinfo {author} {\bibfnamefont {A.~H.}\ \bibnamefont
  {Safavi-Naeini}}, \bibinfo {author} {\bibfnamefont {J.}~\bibnamefont {Chan}},
  \bibinfo {author} {\bibfnamefont {J.~T.}\ \bibnamefont {Hill}}, \bibinfo
  {author} {\bibfnamefont {T.~P.~M.}\ \bibnamefont {Alegre}}, \bibinfo {author}
  {\bibfnamefont {A.}~\bibnamefont {Krause}}, \ and\ \bibinfo {author}
  {\bibfnamefont {O.}~\bibnamefont {Painter}},\ }\href {\doibase
  10.1103/PhysRevLett.108.033602} {\bibfield  {journal} {\bibinfo  {journal}
  {Phys. Rev. Lett.}\ }\textbf {\bibinfo {volume} {108}},\ \bibinfo {pages}
  {033602} (\bibinfo {year} {2012})}\BibitemShut {NoStop}%
\bibitem [{\citenamefont {Zurek}\ \emph {et~al.}(1992)\citenamefont {Zurek},
  \citenamefont {Habib},\ and\ \citenamefont {Paz}}]{zurek1992}%
  \BibitemOpen
  \bibfield  {author} {\bibinfo {author} {\bibfnamefont {W.~H.}\ \bibnamefont
  {Zurek}}, \bibinfo {author} {\bibfnamefont {S.}~\bibnamefont {Habib}}, \ and\
  \bibinfo {author} {\bibfnamefont {J.~P.}\ \bibnamefont {Paz}},\ }\href@noop
  {} {\bibfield  {journal} {\bibinfo  {journal} {Physical Review Letters}\
  }\textbf {\bibinfo {volume} {70}},\ \bibinfo {pages} {1187} (\bibinfo {year}
  {1992})}\BibitemShut {NoStop}%
\bibitem [{\citenamefont {Roucheleau}\ \emph {et~al.}(2010)\citenamefont
  {Roucheleau}, \citenamefont {Ndukum}, \citenamefont {Macklin}, \citenamefont
  {Hertzberg}, \citenamefont {Clerk},\ and\ \citenamefont
  {Schwab}}]{Roucheleau2009}%
  \BibitemOpen
  \bibfield  {author} {\bibinfo {author} {\bibfnamefont {T.}~\bibnamefont
  {Roucheleau}}, \bibinfo {author} {\bibfnamefont {T.}~\bibnamefont {Ndukum}},
  \bibinfo {author} {\bibfnamefont {C.}~\bibnamefont {Macklin}}, \bibinfo
  {author} {\bibfnamefont {J.}~\bibnamefont {Hertzberg}}, \bibinfo {author}
  {\bibfnamefont {A.}~\bibnamefont {Clerk}}, \ and\ \bibinfo {author}
  {\bibfnamefont {K.}~\bibnamefont {Schwab}},\ }\href@noop {} {\bibfield
  {journal} {\bibinfo  {journal} {Nature}\ }\textbf {\bibinfo {volume} {463}},\
  \bibinfo {pages} {72} (\bibinfo {year} {2010})}\BibitemShut {NoStop}%
\bibitem [{\citenamefont {Hertzberg}\ \emph {et~al.}(2009)\citenamefont
  {Hertzberg}, \citenamefont {Rocheleau}, \citenamefont {Ndukum}, \citenamefont
  {Savva}, \citenamefont {Clerk},\ and\ \citenamefont
  {Schwab}}]{hertzberg2009}%
  \BibitemOpen
  \bibfield  {author} {\bibinfo {author} {\bibfnamefont {J.}~\bibnamefont
  {Hertzberg}}, \bibinfo {author} {\bibfnamefont {T.}~\bibnamefont
  {Rocheleau}}, \bibinfo {author} {\bibfnamefont {T.}~\bibnamefont {Ndukum}},
  \bibinfo {author} {\bibfnamefont {M.}~\bibnamefont {Savva}}, \bibinfo
  {author} {\bibfnamefont {A.}~\bibnamefont {Clerk}}, \ and\ \bibinfo {author}
  {\bibfnamefont {K.}~\bibnamefont {Schwab}},\ }\href@noop {} {\bibfield
  {journal} {\bibinfo  {journal} {Nature Physics}\ }\textbf {\bibinfo {volume}
  {6}},\ \bibinfo {pages} {72} (\bibinfo {year} {2009})}\BibitemShut {NoStop}%
\bibitem [{\citenamefont {Murch}\ \emph {et~al.}(2008)\citenamefont {Murch},
  \citenamefont {Moore}, \citenamefont {Gupta},\ and\ \citenamefont
  {Stamper-Kurn}}]{Murch2008}%
  \BibitemOpen
  \bibfield  {author} {\bibinfo {author} {\bibfnamefont {K.}~\bibnamefont
  {Murch}}, \bibinfo {author} {\bibfnamefont {K.}~\bibnamefont {Moore}},
  \bibinfo {author} {\bibfnamefont {S.}~\bibnamefont {Gupta}}, \ and\ \bibinfo
  {author} {\bibfnamefont {D.~M.}\ \bibnamefont {Stamper-Kurn}},\ }\href@noop
  {} {\bibfield  {journal} {\bibinfo  {journal} {Nature Physics}\ }\textbf
  {\bibinfo {volume} {4}},\ \bibinfo {pages} {561} (\bibinfo {year}
  {2008})}\BibitemShut {NoStop}%
\bibitem [{\citenamefont {Purdy}\ \emph {et~al.}(2010)\citenamefont {Purdy},
  \citenamefont {Brooks}, \citenamefont {Botter}, \citenamefont {Brahms},
  \citenamefont {Ma},\ and\ \citenamefont {Stamper-Kurn}}]{Purdy2010}%
  \BibitemOpen
  \bibfield  {author} {\bibinfo {author} {\bibfnamefont {T.~P.}\ \bibnamefont
  {Purdy}}, \bibinfo {author} {\bibfnamefont {D.~W.~C.}\ \bibnamefont
  {Brooks}}, \bibinfo {author} {\bibfnamefont {T.}~\bibnamefont {Botter}},
  \bibinfo {author} {\bibfnamefont {N.}~\bibnamefont {Brahms}}, \bibinfo
  {author} {\bibfnamefont {Z.-Y.}\ \bibnamefont {Ma}}, \ and\ \bibinfo {author}
  {\bibfnamefont {D.~M.}\ \bibnamefont {Stamper-Kurn}},\ }\href {\doibase
  10.1103/PhysRevLett.105.133602} {\bibfield  {journal} {\bibinfo  {journal}
  {Phys. Rev. Lett.}\ }\textbf {\bibinfo {volume} {105}},\ \bibinfo {pages}
  {133602} (\bibinfo {year} {2010})}\BibitemShut {NoStop}%
\bibitem [{\citenamefont {Roach}\ \emph {et~al.}(1964)\citenamefont {Roach},
  \citenamefont {Ketterson},\ and\ \citenamefont {Kuchnir}}]{reppy1964}%
  \BibitemOpen
  \bibfield  {author} {\bibinfo {author} {\bibfnamefont {P.~R.}\ \bibnamefont
  {Roach}}, \bibinfo {author} {\bibfnamefont {J.~B.}\ \bibnamefont
  {Ketterson}}, \ and\ \bibinfo {author} {\bibfnamefont {M.}~\bibnamefont
  {Kuchnir}},\ }\href@noop {} {\bibfield  {journal} {\bibinfo  {journal}
  {Physical Review Letters}\ }\textbf {\bibinfo {volume} {12}},\ \bibinfo
  {pages} {187} (\bibinfo {year} {1964})}\BibitemShut {NoStop}%
\bibitem [{\citenamefont {Larraza}\ and\ \citenamefont
  {Denardo}(1998)}]{Larraza1998}%
  \BibitemOpen
  \bibfield  {author} {\bibinfo {author} {\bibfnamefont {A.}~\bibnamefont
  {Larraza}}\ and\ \bibinfo {author} {\bibfnamefont {B.}~\bibnamefont
  {Denardo}},\ }\href@noop {} {\bibfield  {journal} {\bibinfo  {journal}
  {Physics Letters A}\ }\textbf {\bibinfo {volume} {284}},\ \bibinfo {pages}
  {151} (\bibinfo {year} {1998})}\BibitemShut {NoStop}%
\bibitem [{\citenamefont {Chase}\ and\ \citenamefont
  {Herlin}(1955)}]{chase1955}%
  \BibitemOpen
  \bibfield  {author} {\bibinfo {author} {\bibfnamefont {C.~E.}\ \bibnamefont
  {Chase}}\ and\ \bibinfo {author} {\bibfnamefont {M.~A.}\ \bibnamefont
  {Herlin}},\ }\href@noop {} {\bibfield  {journal} {\bibinfo  {journal}
  {Physical Review}\ }\textbf {\bibinfo {volume} {97}},\ \bibinfo {pages}
  {1447} (\bibinfo {year} {1955})}\BibitemShut {NoStop}%
\bibitem [{\citenamefont {Waters}\ \emph {et~al.}(1967)\citenamefont {Waters},
  \citenamefont {Watmough},\ and\ \citenamefont {Wilks}}]{waters1967}%
  \BibitemOpen
  \bibfield  {author} {\bibinfo {author} {\bibfnamefont {G.~W.}\ \bibnamefont
  {Waters}}, \bibinfo {author} {\bibfnamefont {D.~J.}\ \bibnamefont
  {Watmough}}, \ and\ \bibinfo {author} {\bibfnamefont {J.}~\bibnamefont
  {Wilks}},\ }\href@noop {} {\bibfield  {journal} {\bibinfo  {journal} {Physics
  Letters}\ }\textbf {\bibinfo {volume} {26A}},\ \bibinfo {pages} {12}
  (\bibinfo {year} {1967})}\BibitemShut {NoStop}%
\bibitem [{\citenamefont {Abraham}\ \emph {et~al.}(1968)\citenamefont
  {Abraham}, \citenamefont {Eckstein}, \citenamefont {Ketterson}, \citenamefont
  {Kuchnir},\ and\ \citenamefont {Vignos}}]{abraham1968}%
  \BibitemOpen
  \bibfield  {author} {\bibinfo {author} {\bibfnamefont {B.~M.}\ \bibnamefont
  {Abraham}}, \bibinfo {author} {\bibfnamefont {Y.}~\bibnamefont {Eckstein}},
  \bibinfo {author} {\bibfnamefont {J.~B.}\ \bibnamefont {Ketterson}}, \bibinfo
  {author} {\bibfnamefont {M.}~\bibnamefont {Kuchnir}}, \ and\ \bibinfo
  {author} {\bibfnamefont {J.}~\bibnamefont {Vignos}},\ }\href@noop {}
  {\bibfield  {journal} {\bibinfo  {journal} {Physical Review}\ }\textbf
  {\bibinfo {volume} {181}},\ \bibinfo {pages} {347} (\bibinfo {year}
  {1968})}\BibitemShut {NoStop}%
\bibitem [{\citenamefont {Abraham}\ \emph {et~al.}(1970)\citenamefont
  {Abraham}, \citenamefont {Eckstein}, \citenamefont {Ketterson}, \citenamefont
  {Kuchnir},\ and\ \citenamefont {Roach}}]{Abraham1970}%
  \BibitemOpen
  \bibfield  {author} {\bibinfo {author} {\bibfnamefont {B.~M.}\ \bibnamefont
  {Abraham}}, \bibinfo {author} {\bibfnamefont {Y.}~\bibnamefont {Eckstein}},
  \bibinfo {author} {\bibfnamefont {J.~B.}\ \bibnamefont {Ketterson}}, \bibinfo
  {author} {\bibfnamefont {M.}~\bibnamefont {Kuchnir}}, \ and\ \bibinfo
  {author} {\bibfnamefont {P.~R.}\ \bibnamefont {Roach}},\ }\href {\doibase
  10.1103/PhysRevA.1.250} {\bibfield  {journal} {\bibinfo  {journal} {Phys.
  Rev. A}\ }\textbf {\bibinfo {volume} {1}},\ \bibinfo {pages} {250} (\bibinfo
  {year} {1970})}\BibitemShut {NoStop}%
\bibitem [{\citenamefont {McGuigan}\ \emph {et~al.}(1978)\citenamefont
  {McGuigan}, \citenamefont {Lam}, \citenamefont {Gram}, \citenamefont
  {Hoffman}, \citenamefont {Douglass},\ and\ \citenamefont
  {Gutche}}]{mcguigan1978}%
  \BibitemOpen
  \bibfield  {author} {\bibinfo {author} {\bibfnamefont {D.~F.}\ \bibnamefont
  {McGuigan}}, \bibinfo {author} {\bibfnamefont {C.~C.}\ \bibnamefont {Lam}},
  \bibinfo {author} {\bibfnamefont {R.~Q.}\ \bibnamefont {Gram}}, \bibinfo
  {author} {\bibfnamefont {A.~W.}\ \bibnamefont {Hoffman}}, \bibinfo {author}
  {\bibfnamefont {D.~H.}\ \bibnamefont {Douglass}}, \ and\ \bibinfo {author}
  {\bibfnamefont {H.~W.}\ \bibnamefont {Gutche}},\ }\href@noop {} {\bibfield
  {journal} {\bibinfo  {journal} {Journal of Low Temperature Physics}\ }\textbf
  {\bibinfo {volume} {30}},\ \bibinfo {pages} {621} (\bibinfo {year}
  {1978})}\BibitemShut {NoStop}%
\bibitem [{\citenamefont {Bagdasarov}\ \emph {et~al.}(1977)\citenamefont
  {Bagdasarov}, \citenamefont {Braginsky}, \citenamefont {Mitrofanov},\ and\
  \citenamefont {Shiyan}}]{bagdasarov1977}%
  \BibitemOpen
  \bibfield  {author} {\bibinfo {author} {\bibfnamefont {K.~S.}\ \bibnamefont
  {Bagdasarov}}, \bibinfo {author} {\bibfnamefont {V.~B.}\ \bibnamefont
  {Braginsky}}, \bibinfo {author} {\bibfnamefont {V.~P.}\ \bibnamefont
  {Mitrofanov}}, \ and\ \bibinfo {author} {\bibfnamefont {V.~S.}\ \bibnamefont
  {Shiyan}},\ }\href@noop {} {\bibfield  {journal} {\bibinfo  {journal}
  {Vestnik Moskovskogo Universiteta Seriya Fizika Astronomiya}\ }\textbf
  {\bibinfo {volume} {18}},\ \bibinfo {pages} {98} (\bibinfo {year}
  {1977})}\BibitemShut {NoStop}%
\bibitem [{\citenamefont {Goryachev}\ \emph {et~al.}(2012)\citenamefont
  {Goryachev}, \citenamefont {Creedon}, \citenamefont {Ivanov}, \citenamefont
  {Galliou}, \citenamefont {Bourquin},\ and\ \citenamefont
  {E.}}]{goryachev2012}%
  \BibitemOpen
  \bibfield  {author} {\bibinfo {author} {\bibfnamefont {M.}~\bibnamefont
  {Goryachev}}, \bibinfo {author} {\bibfnamefont {D.~L.}\ \bibnamefont
  {Creedon}}, \bibinfo {author} {\bibfnamefont {E.~N.}\ \bibnamefont {Ivanov}},
  \bibinfo {author} {\bibfnamefont {S.}~\bibnamefont {Galliou}}, \bibinfo
  {author} {\bibfnamefont {R.}~\bibnamefont {Bourquin}}, \ and\ \bibinfo
  {author} {\bibfnamefont {T.~M.}\ \bibnamefont {E.}},\ }\href@noop {}
  {\bibfield  {journal} {\bibinfo  {journal} {Applied Physics Letters}\
  }\textbf {\bibinfo {volume} {100}},\ \bibinfo {pages} {243504} (\bibinfo
  {year} {2012})}\BibitemShut {NoStop}%
\bibitem [{\citenamefont {De~Lorenzo}\ and\ \citenamefont
  {Schwab}()}]{DeLorenzo2013}%
  \BibitemOpen
  \bibfield  {author} {\bibinfo {author} {\bibfnamefont {L.}~\bibnamefont
  {De~Lorenzo}}\ and\ \bibinfo {author} {\bibfnamefont {K.}~\bibnamefont
  {Schwab}},\ }\href@noop {} {\ }\bibinfo {note} {To be published}\BibitemShut
  {NoStop}%
\bibitem [{\citenamefont {Hartung}\ \emph {et~al.}(2006)\citenamefont
  {Hartung}, \citenamefont {Bierwagen}, \citenamefont {Bricker}, \citenamefont
  {Compton}, \citenamefont {Grimm}, \citenamefont {Johnson}, \citenamefont
  {Meidlinger}, \citenamefont {Pendell}, \citenamefont {Popielarski},
  \citenamefont {Saxton},\ and\ \citenamefont {York}}]{hartung2006}%
  \BibitemOpen
  \bibfield  {author} {\bibinfo {author} {\bibfnamefont {W.}~\bibnamefont
  {Hartung}}, \bibinfo {author} {\bibfnamefont {J.}~\bibnamefont {Bierwagen}},
  \bibinfo {author} {\bibfnamefont {S.}~\bibnamefont {Bricker}}, \bibinfo
  {author} {\bibfnamefont {C.}~\bibnamefont {Compton}}, \bibinfo {author}
  {\bibfnamefont {T.}~\bibnamefont {Grimm}}, \bibinfo {author} {\bibfnamefont
  {M.}~\bibnamefont {Johnson}}, \bibinfo {author} {\bibfnamefont
  {D.}~\bibnamefont {Meidlinger}}, \bibinfo {author} {\bibfnamefont
  {D.}~\bibnamefont {Pendell}}, \bibinfo {author} {\bibfnamefont
  {J.}~\bibnamefont {Popielarski}}, \bibinfo {author} {\bibfnamefont
  {L.}~\bibnamefont {Saxton}}, \ and\ \bibinfo {author} {\bibfnamefont {R.~C.}\
  \bibnamefont {York}},\ }in\ \href@noop {} {\emph {\bibinfo {booktitle}
  {Proceedings of LINAC 2006}}}\ (\bibinfo {year} {2006})\ pp.\ \bibinfo
  {pages} {755--757}\BibitemShut {NoStop}%
\bibitem [{\citenamefont {Duffy}\ and\ \citenamefont
  {Umstattd}(1994)}]{duffy1994}%
  \BibitemOpen
  \bibfield  {author} {\bibinfo {author} {\bibfnamefont {J.}~\bibnamefont
  {Duffy}, \bibfnamefont {W.}}\ and\ \bibinfo {author} {\bibnamefont
  {Umstattd}},\ }\href@noop {} {\bibfield  {journal} {\bibinfo  {journal}
  {Journal of Applied Physics}\ }\textbf {\bibinfo {volume} {75}},\ \bibinfo
  {pages} {4489} (\bibinfo {year} {1994})}\BibitemShut {NoStop}%
\bibitem [{ATI()}]{ATIWahChang}%
  \BibitemOpen
  \href@noop {} {}\bibinfo {note} {ATI Wah Chang, 1600 NE Old Salem Road, P.O.
  Box 460, Albany, Oregon 97321}\BibitemShut {NoStop}%
\bibitem [{\citenamefont {Padamsee}(2009)}]{Padamsee2009}%
  \BibitemOpen
  \bibfield  {author} {\bibinfo {author} {\bibfnamefont {H.}~\bibnamefont
  {Padamsee}},\ }\href@noop {} {\emph {\bibinfo {title} {RF Superconductivity:
  Volume II: Science, Technology, and Applications (v.2)}}}\ (\bibinfo
  {publisher} {John Wiley and Sons, Inc.},\ \bibinfo {year} {2009})\BibitemShut
  {NoStop}%
\bibitem [{\citenamefont {Lounasmaa}(1974)}]{Lounasmaa1974}%
  \BibitemOpen
  \bibfield  {author} {\bibinfo {author} {\bibfnamefont {O.}~\bibnamefont
  {Lounasmaa}},\ }\href@noop {} {\emph {\bibinfo {title} {Experimental
  Principles and Methods Below 1K}}}\ (\bibinfo  {publisher} {Academic Press},\
  \bibinfo {address} {New York},\ \bibinfo {year} {1974})\BibitemShut {NoStop}%
\bibitem [{\citenamefont {Sprague}\ and\ \citenamefont
  {Smith}(1998)}]{Sprague1998}%
  \BibitemOpen
  \bibfield  {author} {\bibinfo {author} {\bibfnamefont {D.}~\bibnamefont
  {Sprague}}\ and\ \bibinfo {author} {\bibfnamefont {E.}~\bibnamefont
  {Smith}},\ }\href@noop {} {\bibfield  {journal} {\bibinfo  {journal} {JLTP}\
  }\textbf {\bibinfo {volume} {113}},\ \bibinfo {pages} {975} (\bibinfo {year}
  {1998})}\BibitemShut {NoStop}%
\bibitem [{\citenamefont {McClintock}(1978)}]{mcclintock1978}%
  \BibitemOpen
  \bibfield  {author} {\bibinfo {author} {\bibfnamefont {P.~V.~E.}\
  \bibnamefont {McClintock}},\ }\href@noop {} {\bibfield  {journal} {\bibinfo
  {journal} {Cryogenics}\ }\textbf {\bibinfo {volume} {18}},\ \bibinfo {pages}
  {201} (\bibinfo {year} {1978})}\BibitemShut {NoStop}%
\bibitem [{\citenamefont {Woode}\ \emph {et~al.}(1996)\citenamefont {Woode},
  \citenamefont {Tobar}, \citenamefont {Ivanov},\ and\ \citenamefont
  {Blair}}]{Woode1996}%
  \BibitemOpen
  \bibfield  {author} {\bibinfo {author} {\bibfnamefont {R.}~\bibnamefont
  {Woode}}, \bibinfo {author} {\bibfnamefont {M.}~\bibnamefont {Tobar}},
  \bibinfo {author} {\bibfnamefont {E.}~\bibnamefont {Ivanov}}, \ and\ \bibinfo
  {author} {\bibfnamefont {D.}~\bibnamefont {Blair}},\ }\href@noop {}
  {\bibfield  {journal} {\bibinfo  {journal} {IEEE Trans. Ultrason. Ferroelect.
  Freq. Contr.}\ }\textbf {\bibinfo {volume} {43}},\ \bibinfo {pages} {936}
  (\bibinfo {year} {1996})}\BibitemShut {NoStop}%
\bibitem [{\citenamefont {Abbott}\ and\ \citenamefont
  {et~al.}(2008)}]{Abbott2008}%
  \BibitemOpen
  \bibfield  {author} {\bibinfo {author} {\bibfnamefont {B.}~\bibnamefont
  {Abbott}}\ and\ \bibinfo {author} {\bibnamefont {et~al.}},\ }\href@noop {}
  {\bibfield  {journal} {\bibinfo  {journal} {Astrophysical Journal}\ }\textbf
  {\bibinfo {volume} {683}},\ \bibinfo {pages} {L45} (\bibinfo {year}
  {2008})}\BibitemShut {NoStop}%
\bibitem [{\citenamefont {Press}\ and\ \citenamefont
  {Thorne}(1972)}]{Press1972}%
  \BibitemOpen
  \bibfield  {author} {\bibinfo {author} {\bibfnamefont {W.~H.}\ \bibnamefont
  {Press}}\ and\ \bibinfo {author} {\bibfnamefont {K.~S.}\ \bibnamefont
  {Thorne}},\ }\href@noop {} {\bibfield  {journal} {\bibinfo  {journal}
  {ARA\&A}\ }\textbf {\bibinfo {volume} {10}},\ \bibinfo {pages} {335}
  (\bibinfo {year} {1972})}\BibitemShut {NoStop}%
\bibitem [{\citenamefont {Atkins}\ and\ \citenamefont
  {Stasior}(1953)}]{Atkins1953}%
  \BibitemOpen
  \bibfield  {author} {\bibinfo {author} {\bibfnamefont {K.}~\bibnamefont
  {Atkins}}\ and\ \bibinfo {author} {\bibfnamefont {R.}~\bibnamefont
  {Stasior}},\ }\href@noop {} {\bibfield  {journal} {\bibinfo  {journal} {Can.
  J. Phys.}\ }\textbf {\bibinfo {volume} {41}},\ \bibinfo {pages} {596}
  (\bibinfo {year} {1953})}\BibitemShut {NoStop}%
\end{thebibliography}%

\end{document}